# About "paradox" of negativity of specific heat of the system in thermostat

H. Umirzakov

Institute of Thermophysics, Russia, 630090, Novosibirsk, Lavrentev prospect, 1.

e-mail: cluster125@gmail.com

## Abstract

It is shown that there is in [1] no "proof" of negativity of specific heat of the system placed in thermostat. It is proved that for the system of particles placed in the thermostat and interacting with each other via uniform potential energy the total energy is linear function of the temperature so the isobaric heat capacity is constant along the line of the zero pressure if the latter exists and the isobaric heat capacity is negative for value of the degree of uniformity of the potential energy negative and more than -2 and it is non-negative if otherwise.

The "proof" of the "paradox" of negativity of the isochoric heat capacity (specific heat) of the system of particles in thermostat was published in [1]. But it is strongly proved that the specific heat is positive for both the classical [2] and quantum cases [3]. In this paper we show that there is no "proof" of the negativity of the specific heat in [1].

Let us first repeat the way of the "proof" from [1].

According to the virial theorem [2,4] if $N$ classical particles are placed in closed container with volume $V$, the container is placed in thermostat with the temperature $T$ and the potential $W(\vec{r}_i, i=1..N)$ of interaction of particles with each other ($\vec{r}_i$ is the vector of coordinates of $i$-th particle) obeys the equality

$$W(\lambda \cdot \vec{r}_i, i=1..N) = \lambda^n \cdot W(\vec{r}_i, i=1..N),$$

for any value of $\lambda$ and the velocities of the particles does not diverge than there is the relation

$$2K(T,V) - n \cdot U(T,V) = 3p(T,V) \cdot V,$$

where $K(T,V)$ and $U(T,V)$ are the averaged values of the total kinetic energy of particles and the potential energy $W$ over large time, $p(T,V)$ is the pressure, $n$ is the degree of uniformity of the interaction potential.

For classical particles it is true $K(T,V) = 3NkT/2$ [2,4], so

$$3NkT - n \cdot U(T,V) = 3p(T,V) \cdot V, \qquad (1)$$
$$E(T,V) = 3nkT/2 + U(T,V), \qquad (2)$$

where $k$ is Boltzmann's constant, $E(T,V)$ is the averaged total energy of the particles.

Let us suppose that some $V_0(T)$ exists, so

$$p(T,V)\big|_{V=V_0(T)} = 0. \qquad (3)$$

$V_0(T)$ is the line of zero pressure on the (temperature, volume)- plane. We have from (1)-(2)

$$U(T,V)\big|_{V=V_0(T)} = 3NkT/n, \tag{4}$$

$$E(T,V)\big|_{V=V_0(T)} = 3NkT \cdot (1/2 + 1/n), \tag{5}$$

$$TdS(T,V) = dE(T,V) + p(T,V)dV, \tag{6}$$

$$C_P(T,V(T,p)) = T\left(\frac{\partial S(T,V(T,p))}{\partial T}\right)_P = \left(\frac{\partial E(T,V(T,p))}{\partial T}\right)_P + p \cdot \left(\frac{\partial V(T,p)}{\partial T}\right)_P. \tag{7}$$

Using $E(T,V(T,p))\big|_{p=0} = E(T,V)\big|_{V=V_0(T)}$ we have from (7)

$$C_P(T,p)\big|_{p=0} = \left(\frac{\partial E(T,V(T,p))}{\partial T}\right)_P\bigg|_{p=0} = \frac{dE(T,V)\big|_{V=V_0(T)}}{dT}. \tag{8}$$

Eqs. (5) and (8) give

$$C_P(T,V(T,p))\big|_{p=0} = C_P(T,V)\big|_{V_0(T)} = 3NkT \cdot (1/2 + 1/n), \tag{9}$$

so $C_P(T,V(T,p))\big|_{p=0} < 0$ if the degree of uniformity of the interaction potential obeys the inequalities $0 > n > -2$.

By comparison of above consideration with that of [1] we conclude that there is no proof of negativity of isochoric heat capacity (specific heat) $C_V$ along line of zero compressibility in [1].

The Eq. (9) can be obtained using partition function $Z(T,V)$ of the system under consideration. The partition function if it exists (if it does not diverge) is equal to [4]

$$Z(T,V) = T^{3N \cdot (1/2 + 1/n)} \cdot \exp[f(V \cdot T^{-3/n})]. \tag{10}$$

Using exact relations (5), (10) and

$$F(T,V) = -KT \ln Z(T,V),$$
$$p(T,V) = -(\partial F(T,V)/\partial V)_T,$$
$$S(T,V) = -(\partial F(T,V)/\partial T)_V,$$
$$E(T,V) = F(T,V) + T \cdot S(T,V)$$

one can see that

$$p(T,V) = kT^{1-3/n} \cdot df(x)/dx\big|_{x=VT^{-3/n}}, \tag{11}$$

$$E(T,V) = 3NkT \cdot (1/2 + 1/n) - 3VkT^{1-3/n}/n \cdot df(x)/dx\big|_{x=VT^{-3/n}}. \tag{12}$$

From (3), (11) and (12) we obtain

$$df(x)/dx\big|_{x=V_0(T) \cdot T^{-3/n}} = 0 \tag{13}$$

and Eq. (5) is valid. The relations (5) and (8) give us the relation (9). From (13) we have $f(V_0(T) \cdot T^{-3/n}) = const$, therefore

$$V_0(T) = aT^{3/n}. \tag{14}$$

Using (11), (13) and exact thermodynamic relation [2,4]

$$C_P - C_V = -T[(\partial p/\partial T)_V]^2/(\partial p/\partial V)_T$$

one can obtain

$$C_P(T,V)|_{V=V_0(T)} = C_P(T,V(T,p))|_{p=0} = C_V(T,V)|_{V=V_0(T)} - 9k/n^2 \cdot [x^2 d^2 f(x)/dx^2]|_{x=V_0(T)T^{-3/n}}, \tag{15}$$

so in general case of $[x^2 d^2 f(x)/dx^2]|_{x=V_0(T)T^{-3/n}} \neq 0$ the isochoric heat capacity is not equal to the isobaric one along line of zero pressure.

So we have proved that if the line $V_0(T)$ of zero pressure (3) exists than: a) the isobaric heat capacity along this line is constant; b) it is defined by (9); c) it is equal to the heat capacity of $3N$ harmonic oscillators if $n=2$; d) it is equal to the heat capacity of the ideal gas of $N$ particles if $n >> 2$ or $n << -2$; e) it is negative if the degree of uniformity of the interaction potential obeys the inequalities $0 > n > -2$; f) the total (internal) energy along line of zero pressure (see (5)) is equal to the one of $3N$ harmonic oscillators if $n=2$; g) the total energy along line of zero pressure is equal to the one of the ideal gas of $N$ particles if $|n| >> 2$; h) the line of zero pressure weekly depends on temperature if $|n| >> 2$ (see (14)), so the line of zero compressibility is close to the line of and therefore one can expect that the difference between isochoric and isobaric heat capacities may be small in comparison with the heat capacities.

From (9) and (15) one can have

$$\left|1 - C_V(T,V)|_{V=V_0(T)} / C_P(T,V)|_{V=V_0(T)}\right| = 6/|n \cdot (n+2)| \cdot \left|[x^2 d^2 f(x)/dx^2]|_{x=V_0(T)T^{-3/n}}\right|/N,$$

so if $\left|[x^2 d^2 f(x)/dx^2]|_{x=V_0(T)T^{-3/n}}\right|/N$ is constant or decreases with increasing of absolute value of $n$ and $|n| >> \sqrt{6 \cdot \left|[x^2 d^2 f(x)/dx^2]|_{x=V_0(T)T^{-3/n}}\right|/N}$ then the difference between isochoric and isobaric heat capacities is small in comparison with the heat capacities.

From e) we conclude that if the partition function exists (it is not equal to infinity) for the system of gravitating particles and the line of zero pressure exist than the isobaric heat capacity of the system is constant and negative along the line. The same is true for systems: 1) with coulomb interaction between particles; 2) with coulomb and gravitational interactions between particles. Note that the partition function exists for system with coulomb repulsion.